\RequirePackage{fix-cm}
\documentclass[smallextended]{svjour3}       
\smartqed  
\usepackage{graphicx}
\journalname{Eur. Phys. J. C}
\usepackage{epsfig}
\usepackage{amsmath}
\usepackage{amsfonts,amssymb}
\usepackage{color}

  \usepackage{slashed}
  \usepackage{url}
\usepackage{bm}
\usepackage{verbatim}
\usepackage{float}

\begin{document}

\title{Correction to Nambu-Jona-Lasinio model from QCD at the next-to-leading order}

\author{Marco Frasca         
}


\institute{Marco Frasca \at
              Via Erasmo Gattamelata, 3 \\
							00176 Rome (Italy)
              \email{marcofrasca@mclink.it}           
}

\date{Received: date / Accepted: date}

\maketitle

\begin{abstract}
We evaluate the next-to-leading order correction to the Nambu-Jona-Lasinio model starting from quantum chromodynamics. We show that a systematic expansion exists
, starting from a given set of exact classical solutions, 
so
that higher order corrections could in principle be computed at any order. In this way, we are able to fix the constants of the Nambu-Jona-Lasinio model from quantum chromodynamics and analyze the behavior of strong interactions at low energies. The technique is to expand in powers of currents of the generating functional. We apply it to a simple Yukawa model with self-interaction showing how this has a Nambu-Jona-Lasinio model and its higher order corrections as a low-energy limit. The same is shown to happen for quantum chromodynamics in the chiral limit with two quarks. We prove stability of the NJL model so obtained. Then, we prove that the correction term we obtained does not change the critical temperature of the chiral transition of the Nambu-Jona-Lasinio model at zero chemical potential.
\end{abstract}

\section{Introduction}

Existence of a mass gap for a Yang-Mills theories received a confirmation from recent lattice studies. Strong evidence has been obtained both in the spectrum \cite{Lucini:2004my,Chen:2005mg} and for the gluon propagator \cite{Bogolubsky:2007ud,Cucchieri:2007md,Oliveira:2007px}. On the other side, theoretical analysis have confirmed this finding \cite{Cornwall:1981zr,Cornwall:2010bk,Dudal:2008sp,Tissier:2010ts,Tissier:2011ey,Frasca:2007uz,Frasca:2009yp} but none of them reached the status of a rigorous proof. Anyhow, this fundamental result can be used to perform computations in quantum chromodynamics (QCD) at low energies. The reason is that a very good approximation for the gluon propagator in the Landau gauge in the deep infrared is a free massive propagator as shown in the aforementioned references.

In order to get the gluon propagator in a closed form we need to know what the ground state of a Yang-Mills theory is. We select a ground state using known exact solutions of the classical equations of motion of the theory 
\cite{Frasca:2007uz,Frasca:2009yp,Frasca:2009bc}
. These solutions have the property to recover an instanton liquid for the ground state \cite{Frasca:2013kka,Frasca:2011bd} in agreement with the results given in \cite{Schafer:1996wv}. We will support our choice by comparing the gluon propagator we obtain with the lattice data at large volume presented in \cite{Duarte:2016iko}.

Having the gluon propagator in closed form for a given gauge opens up the possibility to perform computations at low energies in QCD both at zero and finite temperature. Indeed, we have shown in this way that a nonlocal Nambu-Jona-Lasinio (nNJL) model describes the low energy phenomenology of hadron physics \cite{Frasca:2008zp,Frasca:2013kka,Frasca:2012iv,Frasca:2012eq,Frasca:2011bd}. In Ref.\cite{Frasca:2011bd}, we obtained the critical temperature at zero chemical potential for the chiral transition. This turns out in close agreement with lattice data \cite{Lucini:2012gg} and with preceding theoretical computations \cite{GomezDumm:2004sr}. For our aims in this paper we always reduce the nNJL model to a local one. Improvements are left for future works but are not needed here.

The question of how to correct the NJL model is not new \cite{Kashiwa:2006rc,Osipov:2006ns,Hiller:2008nu,Gatto:2010qs,Gatto:2010pt}. These authors do an extensive study of the NJL model with both an eight quark interaction term and a 't~Hooft correction and for zero and non-zero temperature as well. The effect of a magnetic field is also accounted for. These corrections are not generally motivated in a situation where the role of the NJL model is just in the framework of a postulated phenomenological model, even if with great success. The aim of this paper is to provide a sound motivation to the existence of such higher order terms in the NJL model deriving them directly from QCD. This is made possible by the gluon propagator yielded in a closed analytical form and by the existence of a mass gap as said above. This means that the NJL model has a role as a low energy limit of QCD and obtaining higher order corrections is needed due exactly to this role. As a first application we prove that the eight quark interaction term does not change the critical temperature for the chiral transition. We will assume the chiral approximation from the start with the u and d quarks having zero mass. The analysis will be limited to a two-quark model. In this way, this will appear as a leading order approximation to more realistic computations. It is essential to point out that because we start from QCD, all the parameters of the NJL model and its higher order corrections are properly fixed.

The paper has the following structure. In Sec.~\ref{sec2} we introduce a Yukawa model with self-interaction and prove how the infrared limit reduces to a NJL model with an eight quark correction term. In Sec.~\ref{sec2a} we show that the classical solutions we consider are indeed a minimum for the action. In Sec.~\ref{sec3} we do the same starting from a two-quark QCD theory in the chiral limit. In Sec.~\ref{sec4} we discuss the gap equation once all the parameters of the model are properly fixed. In Sec.~\ref{sec5} we prove that, for this model, the eight quark interaction term does not change the critical temperature of the chiral transition. Finally, in Sec.~\ref{sec6} conclusions are given.

\section{Scalar field in the infrared limit\label{sec2}}

Let us consider the generating function of the scalar field
\begin{equation}
    Z[j]=\int[d\phi]\exp\left[-i\int d^4x\left(\frac{1}{2}(\partial\phi)^2-\frac{\lambda}{4}\phi^4+j\phi\right)\right].
\end{equation}
One has
\begin{equation}
   W[j] = -i\int d^4x\left(\frac{1}{2}(\partial\phi)^2-\frac{\lambda}{4}\phi^4+j\phi\right)
\end{equation}
and we can think to perform a functional Taylor series on $W[j]$ \cite{Cahill:1985mh}. We have
\begin{eqnarray}
  W[0] &=& -i\int d^4x\left(\frac{1}{2}(\partial\phi_0)^2-\frac{\lambda}{4}\phi_0^4\right) \nonumber \\
	W_1[0] &=& \left.\frac{\delta W[j]}{\delta j(x_1)}\right|_{j=0} = -i\int d^4x\left(\partial\phi_0(x)\cdot\partial\left.\frac{\delta\phi(x)}{\delta j(x_1)}\right|_{j=0}\right. \nonumber \\
	&&\left.-\lambda\phi_0^3(x)\left.\frac{\delta\phi(x)}{\delta j(x_1)}\right|_{j=0}+\delta^4(x-x_1)\phi_0(x)\right) = -i\phi_0(x_1) \nonumber \\
	W_2[0] &=& \left.\frac{\delta^2 W[j]}{\delta j(x_1)\delta j(x_2)}\right|_{j=0} =
	-i\int d^4x\left(
	\partial\left.\frac{\delta\phi(x)}{\delta j(x_1)}\right|_{j=0}
	\cdot\partial\left.\frac{\delta\phi(x)}{\delta j(x_2)}\right|_{j=0}+\partial\phi_0\cdot\partial\left.\frac{\delta^2\phi(x)}{\delta j(x_1)\delta j(x_2)}\right|_{j=0}\right. \nonumber \\
	&&-3\lambda\phi^2_0(x)\left.\frac{\delta\phi(x)}{\delta j(x_1)}\right|_{j=0}\left.\frac{\delta\phi(x)}{\delta j(x_2)}\right|_{j=0}
	-\lambda\phi_0^3(x)\left.\frac{\delta^2\phi(x)}{\delta j(x_1)\delta j(x_2)}\right|_{j=0} \nonumber \\
	&&\left.+\delta^4(x-x_1)\left.\frac{\delta\phi(x)}{\delta j(x_2)}\right|_{j=0}
	\right) = -2i\left.\frac{\delta\phi(x_2)}{\delta j(x_1)}\right|_{j=0}=-2i\Delta(x_2-x_1) 
\end{eqnarray}
where use has been made of the equations
\begin{eqnarray}
\label{eq:conds}
    \partial^2\phi_0+\lambda\phi_0^3&=&0 \nonumber \\
		\partial^2\left.\frac{\delta\phi(x)}{\delta j(x_1)}\right|_{j=0}+3\lambda\phi_0^2(x)\left.\frac{\delta\phi(x)}{\delta j(x_1)}\right|_{j=0}&=&\delta^4(x-x_1)
\end{eqnarray}
and integration by parts. Then, one has finally
\begin{equation}
\label{eq:W1}
   W[j] = W[0]-i\int d^4xj(x)\phi_0(x)-i\int d^4xd^4x_1j(x)\Delta(x-x_1)j(x_1)+O(j^3).
\end{equation}
Now, let us consider the following Yukawa model
\begin{equation}
   L_Y=\bar\psi(i\slashed\partial-g\phi)\psi+\frac{1}{2}(\partial\phi)^2-\frac{\lambda}{4}\phi^4.
\end{equation}
One has
\begin{equation}
   Z_Y[\bar\eta,\eta]=\int [d\bar\psi][d\psi]\left.\exp\left[-i\int d^4x\bar\psi\left(i\slashed\partial-g\frac{\delta}{\delta j}\right)\psi\right]\exp(W[j])\right|_{j=0}
	\exp\left[i\int d^4x\left(\bar\eta\psi+\bar\psi\eta\right)\right]
\end{equation}
and using eq.(\ref{eq:W1}) yields
\begin{eqnarray}
   Z_Y[\bar\eta,\eta]&=&\int [d\bar\psi][d\psi]\exp\left[-i\int d^4x\bar\psi\left(i\slashed\partial-g\phi_0\right)\psi-g^2\int d^4y\Delta(x-y)\bar\psi(x)\psi(x)\bar\psi(y)\psi(y)\right]\times \nonumber \\
	&&\exp\left[i\int d^4x\left(\bar\eta\psi+\bar\psi\eta\right)\right]
\end{eqnarray}
that is a non-local Nambu-Jona-Lasinio(NJL)-like model in close agreement to \cite{Cahill:1985mh}. In order to correct this approximation, we consider the next-to-leading order correction that is
\begin{eqnarray}
   W_3[0] &=& \left.\frac{\delta^3 W[j]}{\delta j(x_1)\delta j(x_2)\delta j(x_3)}\right|_{j=0} = \nonumber \\
	&&-i\int d^4x\left[\partial\left.\frac{\delta^2\phi}{\delta j(x_2)\delta j(x_3)}\right|_{j=0}\partial\left.\frac{\delta\phi}{\delta j(x_1)}\right|_{j=0}
	+\partial\left.\frac{\delta^2\phi}{\delta j(x_1)\delta j(x_3)}\right|_{j=0}\partial\left.\frac{\delta\phi}{\delta j(x_2)}\right|_{j=0}\right. \nonumber \\
	&&+\partial\left.\frac{\delta^2\phi}{\delta j(x_1)\delta j(x_2)}\right|_{j=0}\partial\left.\frac{\delta\phi}{\delta j(x_3)}\right|_{j=0} 
	+\partial\phi_0\partial\left.\frac{\delta^3\phi}{\delta j(x_1)\delta j(x_2)\delta j(x_3)}\right|_{j=0}\nonumber \\
	&&
	-3\lambda\phi_0^2\left.\frac{\delta^2\phi}{\delta j(x_2)\delta j(x_3)}\right|_{j=0}\left.\frac{\delta\phi}{\delta j(x_1)}\right|_{j=0}
	-3\lambda\phi_0^2\left.\frac{\delta^2\phi}{\delta j(x_1)\delta j(x_3)}\right|_{j=0}\left.\frac{\delta\phi}{\delta j(x_2)}\right|_{j=0} \nonumber \\
	&&
	-3\lambda\phi_0^2\left.\frac{\delta^2\phi}{\delta j(x_1)\delta j(x_2)}\right|_{j=0}\left.\frac{\delta\phi}{\delta j(x_3)}\right|_{j=0}
	-6\lambda\phi_0\left.\frac{\delta\phi}{\delta j(x_1)}\right|_{j=0}\left.\frac{\delta\phi}{\delta j(x_2)}\right|_{j=0}\left.\frac{\delta\phi}{\delta j(x_3)}\right|_{j=0} \nonumber \\
	&&-\lambda\phi_0^3\left.\frac{\delta^3\phi}{\delta j(x_1)\delta j(x_2)\delta j(x_3)}\right|_{j=0} \nonumber \\
	&&\left.+\delta^4(x-x_1)\left.\frac{\delta^2\phi}{\delta j(x_2)\delta j(x_3)}\right|_{j=0}
	+\delta^4(x-x_2)\left.\frac{\delta^2\phi}{\delta j(x_1)\delta j(x_3)}\right|_{j=0}
	+\delta^4(x-x_3)\left.\frac{\delta^2\phi}{\delta j(x_1)\delta j(x_2)}\right|_{j=0}
	\right] \nonumber \\
	&&= 6i\lambda\int d^4x\phi_0(x)\Delta(x-x_1)\Delta(x-x_2)\Delta(x-x_3)
\end{eqnarray}
where use has been mad of eq.(\ref{eq:conds}) and integration by parts. A similar computation yields
\begin{equation}
   W_4[0] = \left.\frac{\delta^4 W[j]}{\delta j(x_1)\delta j(x_2)\delta j(x_3)\delta j(x_4)}\right|_{j=0} = 6i\lambda\int d^4x\Delta(x-x_1)\Delta(x-x_2)\Delta(x-x_3)\Delta(x-x_4)
\end{equation}
where use has been made of the equation
\begin{equation}
   \partial^2\left.\frac{\delta^2\phi}{\delta j(x_1)\delta j(x_2)}\right|_{j=0}
	+\lambda\left[6\phi_0(x)\Delta(x-x_1)\Delta(x-x_2)+3\phi_0^2(x)\left.\frac{\delta^2\phi}{\delta j(x_1)\delta j(x_2)}\right|_{j=0}\right]=0.
\end{equation}
The functional becomes
\begin{eqnarray}
\label{eq:W2}
   W[j] &=& W[0]-i\int d^4xj(x)\phi_0(x)-i\int d^4xd^4x_1j(x)\Delta(x-x_1)j(x_1) \nonumber \\
	&&+i\lambda\int d^4xd^4x_1d^4x_2d^4x_3\phi_0(x)\Delta(x-x_1)j(x_1)\Delta(x-x_2)j(x_2)\Delta(x-x_3)j(x_3) \nonumber \\
	&&+\frac{i}{4}\lambda\int d^4xd^4x_1d^4x_2d^4x_3d^4x_4\Delta(x-x_1)j(x_1)\Delta(x-x_2)j(x_2)\Delta(x-x_3)j(x_3)\Delta(x-x_4)j(x_4) \nonumber \\
	&&+O(j^5).
\end{eqnarray}
This permits to compute the next-to-leading order correction to the non-local NJL-like model we obtained above. The model has now the structure
\begin{eqnarray}
    S_{NJL} &=& \int d^4x\left[\bar\psi(x)\left(i\slashed\partial-g\phi_0(x)\right)\psi(x)-g^2\int d^4x_1\Delta(x-x_1)\bar\psi(x)\psi(x)\bar\psi(x_1)\psi(x_1)\right. \\
		&&+g^3\lambda\int d^4x_1d^4x_2d^4x_3\phi_0(x)\Delta(x-x_1)\bar\psi(x_1)\psi(x_1)\Delta(x-x_2)\bar\psi(x_2)\psi(x_2)\Delta(x-x_3)\bar\psi(x_3)\psi(x_3) \nonumber \\
		&&+\frac{\lambda}{4}g^4
		\int d^4x_1d^4x_2d^4x_3d^4x_4\Delta(x-x_1)\bar\psi(x_1)\psi(x_1)\Delta(x-x_2)\bar\psi(x_2)\psi(x_2)\Delta(x-x_3)\bar\psi(x_3)\psi(x_3)\times  \nonumber \\
		&&\left.\Delta(x-x_4)\bar\psi(x_4)\psi(x_4)
		\right] \nonumber
\end{eqnarray}
and we observe that odd powers of the current take a $\phi_0$ contribution. The important conclusion to be drawn is that {\sl if one knows an exact solution to the equation of motion and the corresponding propagator (see eq.(\ref{eq:conds})), the scalar theory is completely solved and the Yukawa model reduces to a non-local NJL model and its higher order corrections.} The latter are easy to compute due to the triviality of the scalar theory.

To fix the parameters of the theory, we now consider an exact solution \cite{Frasca:2013tma} $\phi_0(x) = \mu\left(2/\lambda\right)^\frac{1}{4}{\rm sn}(p\cdot x+\theta,i)$ being {\rm sn} an elliptic Jacobi function and $\mu$ and $\theta$ two integration constants. In order for this solutions to hold, the following dispersion relation applies $p^2=\mu^2\sqrt{\lambda/2}$. We recognize here a free massive solution notwithstanding we started from a massless theory. The corresponding 2-point function can be computed immediately yielding \cite{Frasca:2013tma}
\begin{equation}
\label{eq:green}
   \Delta(p)=\sum_{n=0}^\infty\frac{B_n}{p^2-m_n^2+i\epsilon}
\end{equation}
being
\begin{equation}
   B_n=(2n+1)^2\frac{\pi^3}{4K^3(i)}\frac{e^{-(n+\frac{1}{2})\pi}}{1+e^{-(2n+1)\pi}},
\end{equation}
the mass spectrum is
\begin{equation}
\label{eq:ms}
   m_n=(2n+1)(\pi/2K(i))\left(\lambda/2\right)^{\frac{1}{4}}\mu
\end{equation}
and $K(i)\approx 1.3111028777$ an elliptic integral, consistently with the idea of a strong coupling expansion. This holds provided one fixes the phase $\theta$ in the exact solution to $\theta_m=(4m+1)K(i)$. This identifies an infinite set of scalar field theories with a trivial infrared fixed point in quantum field theory. These solutions appear really interesting as they provide a strong coupling expansion for eq.(\ref{eq:W2}) in inverse powers of $\lambda$. It is very easy to see that the contact limit is obtained by taking $p=0$ in the propagator that provides
\begin{equation}
   \Delta(x-y)\approx -\sum_{n=0}^\infty\frac{B_n}{m_n^2}\delta^4(x-y)=-\lambda^{-\frac{1}{2}}{\cal K}\delta^4(x-y),
\end{equation}
being ${\cal K}$ a constant having the dimensions of inverse mass squared, taking us to the well-known contact interaction of a NJL model. Similarly, we note that $\phi_0(x)=\lambda^{-\frac{1}{4}}\chi(x)$. Our action has now the form
\begin{eqnarray}
    S_{NJL} &=& \int d^4x\left[\bar\psi(x)\left(i\slashed\partial-g\lambda^{-\frac{1}{4}}\chi(x)\right)\psi(x)+g^2{\cal K}\lambda^{-\frac{1}{2}}\bar\psi(x)\psi(x)\bar\psi(x)\psi(x)\right. \\
		&&-g^3{\cal K}^3\lambda^{-\frac{3}{4}}\chi(x)\bar\psi(x)\psi(x)\bar\psi(x)\psi(x)\bar\psi(x)\psi(x) \nonumber \\
		&&\left.+g^4{\cal K}^4\lambda^{-1}\frac{1}{4}\bar\psi(x)\psi(x)\bar\psi(x)\psi(x)\bar\psi(x)\psi(x)\bar\psi(x)\psi(x)\right]. \nonumber
\end{eqnarray}
Now, by assuming the average of the field $\chi(x)$ to be zero, we recover a standard mean field NJL model with a higher order correction
\begin{eqnarray}
    S_{NJL} &=& \int d^4x\left[\bar\psi(x)i\slashed\partial\psi(x)+g^2{\cal K}\lambda^{-\frac{1}{2}}\bar\psi(x)\psi(x)\bar\psi(x)\psi(x)\right. \\
		&&\left.+g^4{\cal K}^4\lambda^{-1}\frac{1}{4}\bar\psi(x)\psi(x)\bar\psi(x)\psi(x)\bar\psi(x)\psi(x)\bar\psi(x)\psi(x)\right]. \nonumber
\end{eqnarray}


\section{The question of the minimum\label{sec2a}}

One could ask if the exact solutions we started from to build the expansion for the scalar field are a real minimum for the functional of the field. This, in view of the mapping theorem proven in \cite{Frasca:2007uz,Frasca:2009yp}, this will immediately apply to the Yang-Mills equations we exploit in the following sections. That this is so can be seen in the following way. Firstly, let us see how the idea works for the free field. One has
\begin{equation}
    {\cal L}[\phi] = \int d^4x\left[\frac{1}{2}(\partial\phi)^2-\frac{1}{2}m^2\phi^2\right].
\end{equation}
For a given classical solution $\phi_0$, we can take a functional Taylor series of this as
\begin{equation}
    {\cal L} = {\cal L}[\phi_0]+\int d^4x'\left.\frac{\delta{\cal L}}{\delta\phi(x')}\right|_{\phi=\phi_0}\phi(x')
		+\frac{1}{2}\int d^4x'd^4x''\left.\frac{\delta^2{\cal L}}{\delta\phi(x')\delta\phi(x'')}\right|_{\phi=\phi_0}\phi(x')\phi(x'')+\ldots
\end{equation}
that becomes
\begin{equation}
    {\cal L} = {\cal L}[\phi_0]
		-\frac{1}{2}\int d^4x'd^4x\left[\partial^2\delta^4(x-x')+m^2\delta^4(x-x')\right]\phi(x')\phi(x)+\ldots,    
\end{equation}
where use has been made of the condition $\left.\frac{\delta{\cal L}}{\delta\phi(x')}\right|_{\phi=\phi_0}=0$, due to motion equation. We can now prove that the term originating from the second functional derivative is indeed strictly positive off-shell. Let us consider the eigenvalue problem
\begin{equation}
   \partial^2\chi_n+m^2\chi_n=\lambda_n\chi_n.
\end{equation}
Then, one has $\delta^4(x-x')=\sum_n\chi_n(x)\chi_n(x')$. It is not difficult to see that $\lambda_n=\lambda(p)=-(p^2-m^2)\le 0$ as the theory has a lower bound to energy. This means that
\begin{equation}
    {\cal L} = {\cal L}[\phi_0]+\frac{1}{2}\int d^4x'd^4x\sum_p(p^2-m^2)\chi_p(x)\chi_p(x')\phi(x')\phi(x)+\ldots.    
\end{equation}
But,
\begin{equation}
   c(p)=\int d^4x\chi_p(x)\phi(x)
\end{equation}
are the coefficients of the Fourier series for the field in terms of the eigenfunctions $\chi_p$ and then
\begin{equation}
    {\cal L} = {\cal L}[\phi_0]+\frac{1}{2}\sum_p(p^2-m^2)c^2(p)+\ldots={\cal L}[\phi_0]+\frac{1}{2}\sum_p(p^2-m^2)c^2(p)+\ldots.    
\end{equation}
The second term on the lhs must be positive definite for physical reasons and the classical solution is a minimum of the functional. For this it is enough to assume a lower bound on the spectrum for $p^2=m^2$ that grants the positivity of the energy. The zero mode is expected due to the translational invariance of the theory.

We can use this strategy to show that the classical solutions we use in our analysis are indeed stable solutions. Now it is
\begin{equation}
    {\cal L}[\phi] = \int d^4x\left[\frac{1}{2}(\partial\phi)^2-\frac{\lambda}{4}\phi^4\right].
\end{equation}
Then,
\begin{equation}
    {\cal L} = {\cal L}[\phi_0]
		-\frac{1}{2}\int d^4x'd^4x\left[\partial^2\delta^4(x-x')+3\lambda\phi_0^2(x)\delta^4(x-x')\right]\phi(x')\phi(x)+\ldots    
\end{equation}
being now $\phi_0$ given by a solution to $\partial^2\phi_0+\lambda\phi_0^3=0$ (see \cite{Frasca:2009bc}). We introduce the set of eigenfunctions
\begin{equation}
    \partial^2\varphi_n+3\lambda\phi_0^2(x)\varphi_n=\epsilon_n\varphi_n.
\end{equation}
So, let us consider $phi_0=\mu(2/\lambda)^\frac{1}{4}{\rm sn}(p\cdot x,-1)$ being sn a Jacobi elliptical function, then the eigenfunctions take the form
\begin{equation}
     \varphi_\mu=C\cdot{\rm sn}(p\cdot x,-1){\rm cn}(p\cdot x,-1)
\end{equation}
with eigenvalues $\epsilon(\mu)=-3\mu^2\sqrt{\lambda/2}\le 0$ with $\mu$ running from 0 to infinity. This holds on-shell and shows that the argument that works for the free field case holds also here and our classical solutions are a minimum for the action functional, provided we work on-shell. We see again the zero mode showing that translational invariance applies for the theory.

This applies directly to Yang-Mills theory as we have shown a mapping theorem between the solutions of the scalar field theory and those of Yang-Mills theory\cite{Frasca:2007uz,Frasca:2009yp,Frasca:2009bc}. This theorem grants that, just in the Landau gauge, the mapping is exact.\footnote{The correctness of this theorem was agreed with T.~Tao after that a problem in the proof was properly fixed (see \url{http://wiki.math.toronto.edu/DispersiveWiki/index.php/Talk:Yang-Mills_equations).}} 

\section{QCD in the infrared limit\label{sec3}}

\begin{figure}[H]
  \includegraphics{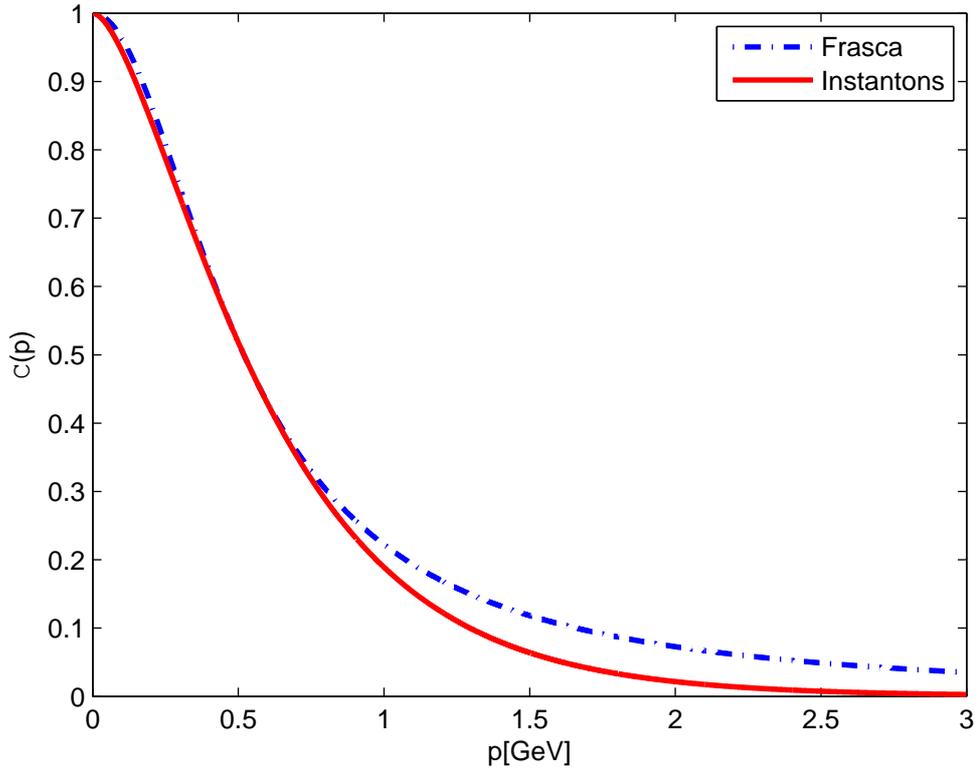}
  \caption{Comparison of our form factor with that provided in \cite{Schafer:1996wv} 
  for $\sqrt{\sigma}=0.417\ GeV$ and $d^{-1}=0.58\ GeV$.\label{fig:ff}}
\end{figure}

Differently from the Yukawa model, instead than a scalar field we have to cope with a Yang-Mills field. So, we have to start from the functional
\begin{eqnarray}
   W_{YM} &=& -\frac{1}{4}\int d^4x\left[(\partial_\mu A_\nu^a-\partial_\nu A_\mu^a)(\partial^\mu A^{a\nu}-\partial^\nu A^{a\mu})\right. \nonumber \\
	        && +2gf^{abc}(\partial_\mu A_\nu^a-\partial_\nu A_\mu^a)A^{\mu b}A^{\nu c} \nonumber \\
	        && +g^2f^{abc}f^{ade}A^{\mu b}A^{\nu c}A^d_\mu A^e_\nu \nonumber \\
					&& +\left.\frac{1}{\alpha}(\partial A)^2\right] \nonumber \\
					&& -\int d^4x(\bar c^a\partial_\mu\partial^\mu c^a+g\bar c^a f^{abc}\partial_\mu A^{b\mu}c^c) \nonumber \\
					&& +\int d^4x A_\mu^aj^{\mu a}+\int d^4x(\bar c^a\epsilon^a+\bar\epsilon^ac^a)
\end{eqnarray}
being $c^a,\ \bar c^a$ the ghost fields. The infrared behavior of the Yang-Mills field resembles very near the behavior of the quartic scalar field. Indeed, classical solutions can be mapped between these two theories \cite{Frasca:2009yp,Frasca:2015yva}. From lattice computations \cite{Bogolubsky:2007ud,Cucchieri:2007md,Oliveira:2007px}, we know that, in the Landau gauge, the ghost field decouples and behaves as free, the gluon propagator in the deep infrared maps very well a free massive propagator and the running coupling seems to go to zero \cite{Bogolubsky:2009dc} marking possibly a trivial theory in this limit. This scenario appears to be confirmed by a renormalization group analysis \cite{Weber:2011nw}. So, the gluon propagator in the Landau gauge takes the form
\begin{equation}
   \Delta_{\mu\nu}^{ab}(p)=\delta_{ab}\left(\eta_{\mu\nu}-\frac{p_\mu p_\nu}{p^2}\right)\Delta(p)
\end{equation}
and, assuming an almost trivial theory in the infrared limit, the above functional can be expanded in currents as
\begin{eqnarray}
   W_{YM}[j,\epsilon,\bar\epsilon] &=& \int d^4x\bar\epsilon(x)G(x)\epsilon(x)+\int d^4x\Phi_\mu^a(x)j^{\mu a}(x)
	-\int d^4x_1d^4x_2j^{\mu a}(x_1)\Delta_{\mu\nu}^{ab}(x_1-x_2)j^{\nu b}(x_2) \nonumber \\
	&&+\int d^4xd^4x_1d^4x_2d^4x_3\Phi^{\mu a}(x)\Delta_{\mu\nu}^{ab}(x-x_1)j^{\nu b}(x_1)\Delta_{\kappa\lambda}^{cd}(x-x_2)j^{\kappa c}(x_2)\Delta^{\lambda de}_\rho(x-x_3)j^{\rho e}(x_3)
	\nonumber \\
	&&+\int
	d^4xd^4x_1d^4x_2d^4x_3d^4x_4\Delta_{\mu\nu}^{ab}(x-x_1)j^{\mu a}(x_1)\Delta_{\lambda}^{\nu bc}(x-x_2)j^{\lambda c}(x_2)\times \nonumber \\
	&&\Delta^{de}_{\kappa\rho}(x-x_3)j^{\kappa d}(x_3)
	\Delta^{\rho ef}_{\theta}(x-x_4)j^{\theta f}(x_4) + O(j^5).
\end{eqnarray}
From this equation, it is easy to obtain the non-local NJL and its next-to-leading order correction with the substitution $j^{\mu a}(x)\rightarrow -g\sum_q\bar q(x)\frac{\lambda^a}{2}\gamma^\mu q(x)$ being $q(x)$ the quark field and $\lambda^a$ SU(3) matrices, fixing the gauge to Landau and using current conservation \cite{Frasca:2011bd,Frasca:2013kka}. We will get
\begin{eqnarray}
   W_{YM}[\Phi,q,\bar q] &=& -g\int d^4x\Phi_\mu^a(x)\sum_q\bar q(x)\frac{\lambda^a}{2}\gamma^\mu q(x) \nonumber \\
	&&-g^2\int d^4x_1d^4x_2\sum_q\bar q(x_1)\frac{\lambda^a}{2}\gamma^\mu q(x_1)\Delta(x_1-x_2)\sum_{q'}\bar q'(x_2)\frac{\lambda^a}{2}\gamma_\mu q'(x_2) \nonumber \\
	&&-g^3\int d^4xd^4x_1d^4x_2d^4x_3\Phi^{\mu a}(x)\Delta(x-x_1)\sum_q\bar q(x_1)\frac{\lambda^a}{2}\gamma_\mu q(x_1)\times \nonumber \\
	&&\Delta(x-x_2)\sum_{q'}\bar q'(x_2)\frac{\lambda^b}{2}\gamma^\nu q'(x_2)\Delta(x-x_3)
	\sum_{q''}\bar q''(x_3)\frac{\lambda^b}{2}\gamma_\nu q''(x_3) \nonumber \\
	&&+g^4\int
	d^4xd^4x_1d^4x_2d^4x_3d^4x_4\Delta(x-x_1)\sum_q\bar q(x_1)\frac{\lambda^a}{2}\gamma^\mu q(x_1)
	\Delta(x-x_2)\sum_{q'}\bar q'(x_2)\frac{\lambda^a}{2}\gamma_\mu q'(x_2)\times \nonumber \\
	&&\Delta(x-x_3)\sum_{q''}\bar q''(x_3)\frac{\lambda^b}{2}\gamma^\nu q''(x_3)
	\Delta(x-x_4)\sum_{q'''}\bar q'''(x_4)\frac{\lambda^b}{2}\gamma_\nu q'''(x_4)
\end{eqnarray}
where we omitted the ghost contribution as, in the infrared limit, this decouples. This is a completely non-local NJL model and its higher order corrections properly derived from QCD. We can omit the term with the $\Phi$ solution provided it averages to zero yielding a mean field approximation. This yields the result
\begin{eqnarray}
   W_{NJL}[q,\bar q] &=& -g^2\int d^4x_1d^4x_2\sum_q\bar q(x_1)\frac{\lambda^a}{2}\gamma^\mu q(x_1)\Delta(x_1-x_2)\sum_{q'}\bar q'(x_2)\frac{\lambda^a}{2}\gamma_\mu q'(x_2) \nonumber \\
	&&+g^4\int
	d^4xd^4x_1d^4x_2d^4x_3d^4x_4\Delta(x-x_1)\sum_q\bar q(x_1)\frac{\lambda^a}{2}\gamma^\mu q(x_1)
	\Delta(x-x_2)\sum_{q'}\bar q'(x_2)\frac{\lambda^a}{2}\gamma_\mu q'(x_2)\times \nonumber \\
	&&\Delta(x-x_3)\sum_{q''}\bar q''(x_3)\frac{\lambda^b}{2}\gamma^\nu q''(x_3)
	\Delta(x-x_4)\sum_{q'''}\bar q'''(x_4)\frac{\lambda^b}{2}\gamma_\nu q'''(x_4)
\end{eqnarray}
Then, we are able to recover the non-local Nambu-Jona-Lasinio model in the way yielded in \cite{Frasca:2011bd,Frasca:2013kka} properly corrected, directly from QCD, provided the form factor is
\begin{equation}
\label{eq:Gp}
      {\cal G}(p)=-\frac{1}{2}g^2\Delta(p)=-\frac{1}{2}g^2\sum_{n=0}^\infty\frac{B_n}{p^2-(2n+1)^2(\pi/2K(i))^2\tilde\sigma+i\epsilon}
      =\frac{G}{2}{\cal C}(p)
\end{equation}
being $B_n$ obtained from eq.(\ref{eq:green}), $\tilde\sigma$ a constant having the dimensions of a squared mass, ${\cal C}(0)=1$ and $2{\cal G}(0)=G$ the standard Nambu-Jona-Lasinio coupling, fixing in this way the value of $G$ through the gluon propagator. We just note that $G=\frac{g^2}{(\pi/2K(-1))^2\tilde\sigma}\sum_{n=0}{\infty}\frac{B_n}{(2n+1)^2}\approx 0.7854(g^2/\tilde\sigma)$. In Fig.\ref{fig:ff}, we compare this form factor both with the one from an instanton liquid \cite{Schafer:1996wv} that is
\begin{equation}
\mathcal{C}_I(\xi)=4\pi^2 d^2\left\{\xi\dfrac{d}{d\xi}\big[I_0(\xi)K_0(\xi)-I_1(\xi)K_1(\xi)\big]\right\}^2\qquad\text{with } \xi=\frac{|p| d}{2} 
\end{equation}
being $I_n$ and $K_n$ Bessel functions. In the following we normalize this function to be 1 at zero momenta dividing it by ${\cal C}_I(0)$.
The result is strikingly good for the latter showing how consistently our technique represents Yang-Mills theory through instantons. 
One can object this choice of the gluon propagator, that anyway performs so well with respect to a ground state being an instanton liquid, by claiming that the Yang-Mills functional should select exactly it. Of course, one can do quantum field theory around any kind of classical solution representing in this way the ground state of the theory. Experiments should say the last word about. For Yang-Mills theories that exist in nature only interacting with other fields, the last word is in lattice computations. So, we compare our propagator with lattice data at $128^4$ points, a very large volume, and $\beta=6$ obtained in \cite{Duarte:2016iko}. The result is provided in the following figure.
\begin{figure}[H]
  \includegraphics{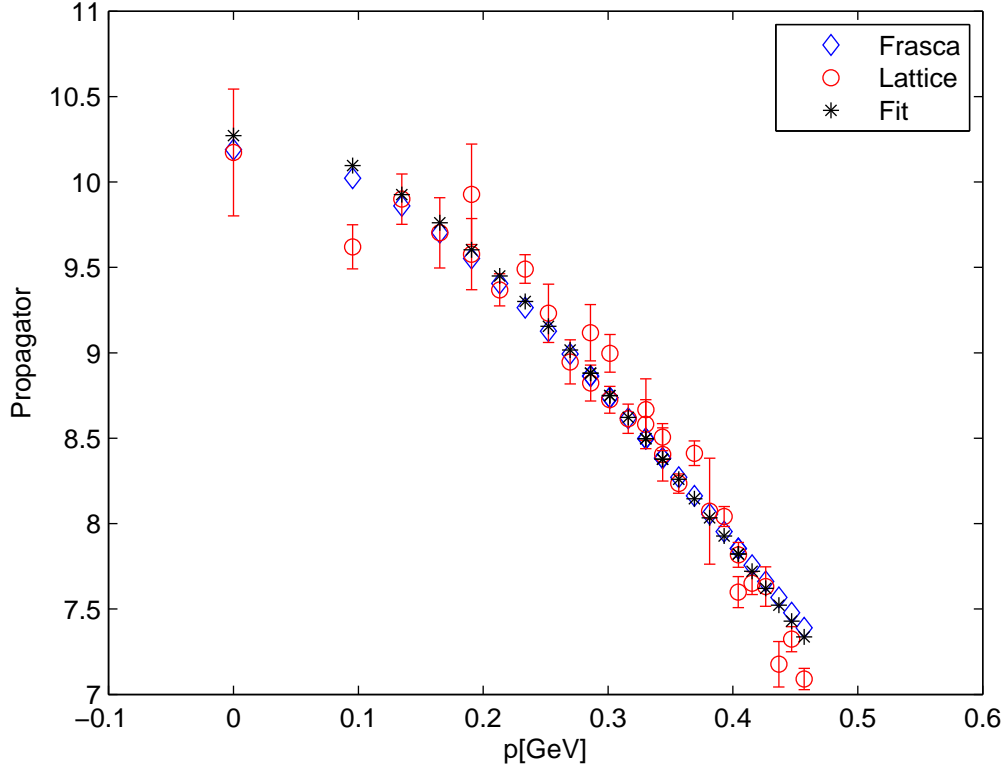}
  \caption{Comparison in the deep infrared region of our propagator (diamonds with error bars) with that obtained with lattice computations in \cite{Duarte:2016iko} 
  for a volume of $128^4$ points and $\beta=6$ (circles). A fit with a Yukawa propagator is also provided (asterisks). \label{fig:fit}}
\end{figure}
The agreement is exceedingly good confirming that our choice of the ground state is a sound one.

So,
if we limit our analysis to u and d quarks so that, $\psi=(u,d)$, we can rewrite our Lagrangian as
\begin{eqnarray}
   W_{NJL}[q,\bar q] &=& \int d^4x_1d^4x_2{\cal G}(x_1-x_2)\left[\bar\psi(x_1)\gamma^\mu\psi(x_1)\bar\psi(x_2)\gamma_\mu\psi(x_2)+\bar\psi(x_1)\gamma^\mu{\bm\tau}\psi(x_1)\bar\psi(x_2)\gamma_\mu{\bm\tau}\psi(x_2)\right] \nonumber \\
	&&+g^{-4}\int
	d^4xd^4x_1d^4x_2d^4x_3d^4x_4{\cal G}(x-x_1){\cal G}(x-x_2){\cal G}(x-x_3){\cal G}(x-x_4)\times \nonumber \\
	&&\left[\bar\psi(x_1)\gamma^\mu\psi(x_1)\bar\psi(x_2)\gamma_\mu\psi(x_2)+\bar\psi(x_1)\gamma^\mu{\bm\tau}\psi(x_1)\bar\psi(x_2)\gamma_\mu{\bm\tau}\psi(x_2)\right]\times \\
	&&\left[\bar\psi(x_3)\gamma^\mu\psi(x_3)\bar\psi(x_4)\gamma_\mu\psi(x_4)+\bar\psi(x_3)\gamma^\mu{\bm\tau}\psi(x_3)\bar\psi(x_4)\gamma_\mu{\bm\tau}\psi(x_4)\right].
\end{eqnarray}
Finally, collapsing to the local limit we get
\begin{eqnarray}
   W_{NJL}[q,\bar q] &=& \frac{G}{2}\int d^4x\left[\bar\psi(x)\gamma^\mu\psi(x)\bar\psi(x)\gamma_\mu\psi(x)+\bar\psi(x)\gamma^\mu{\bm\tau}\psi(x)\bar\psi(x)\gamma_\mu{\bm\tau}\psi(x)\right] \nonumber \\
	&&+G_8\int d^4x\left[\bar\psi(x)\gamma^\mu\psi(x)\bar\psi(x)\gamma_\mu\psi(x)+\bar\psi(x)\gamma^\mu{\bm\tau}\psi(x)\bar\psi(x)\gamma_\mu{\bm\tau}\psi(x)\right]\times \nonumber \\
	&&\left[\bar\psi(x)\gamma^\mu\psi(x)\bar\psi(x)\gamma_\mu\psi(x)+\bar\psi(x)\gamma^\mu{\bm\tau}\psi(x)\bar\psi(x)\gamma_\mu{\bm\tau}\psi(x)\right]
\end{eqnarray}
being 
\begin{equation}
   {\cal G}_8(x-x_1,x-x_2,x-x_3,x-x_4)=g^{-4}{\cal G}(x-x_1){\cal G}(x-x_2){\cal G}(x-x_3){\cal G}(x-x_4)
\end{equation}
that yields
\begin{equation}
   G_8 = g^{-4}\left[{\cal G}(0)\right]^4=\frac{g^4}{(\pi/2K(-1))^8{\tilde\sigma}^4}
	\left[\sum_{n=0}^{\infty}\frac{B_n}{(2n+1)^2}\right]^4\approx 0.38\frac{g^4}{{\tilde\sigma}^4}.
\end{equation}
So, as expected, $G_8$ has the dimensions of the inverse of the eighth power of a mass.

We see that QCD is able to fix all the parameters of the low-energy NJL model and we obtained in this way the next-to-leading order correction that is an eight quark interaction term. This has been extensively analysed in literature and, in the following we revise the main results with the new parameters we were able to get.

\section{Gap equation\label{sec4}}

Gap equation for an extended NJL model has been obtained in \cite{Hiller:2008nu,Osipov:2006ns}. We map our model on theirs. These authors consider a 't~Hooft term with coupling $\kappa$, that we do not have in the averaged model, and two terms for the eight quark interaction with coupling $g_1$ and $g_2$. To map the two models we have just to put $\kappa=0$ and $g_2=0$ while is $g_1=G_8/8$. The quartic term remains untouched. So, the effective potential \cite{Osipov:2006ns} is
\begin{eqnarray}
    V(M)&=&\frac{h^2}{16}\left(12G+\frac{27}{16}G_8h^2\right) \nonumber \\
		&&-\frac{3N}{16\pi^2}\left[M^2J_0(M^2)+\Lambda^4\ln\left(1+\frac{M^2}{\Lambda^2}\right)\right]
\end{eqnarray}
being $M$ the effective mass quark to be computed, $\Lambda$ the cut-off needed to regularize the NJL model, $N$ the number of colors, and $h=h(M)$ is obtained by solving the equation
\begin{equation}
    M+Gh+\frac{3}{32}G_8h^3=0.
\end{equation}
$J_0$ is one of the integrals of the NJL model that has the value
\begin{equation}
\label{eq:J_0}
    J_0(M^2)=16\pi^2i\int_\Lambda\frac{d^4p}{(2\pi)^4}\frac{1}{p^2-M^2}=\Lambda^2-M^2\ln\left(1+\frac{M^2}{\Lambda^2}\right).
\end{equation}
where a cut-off regularization is intended and this cut-off is $\Lambda$. In agreement with \cite{Osipov:2006ns}, this model is stable as the third degree equation for $h$ has just one real solution. We can evaluate it iteratively assuming that $G_8h^3$ is just a small correction to our model. One has
\begin{equation}
   h\approx -\frac{M}{G}+\frac{3}{32}\frac{G_8M^3}{G^3}.
\end{equation}
Anyhow, we prefer to solve it numerically. We plot the effective potential assuming $G\approx 9.37\ GeV^{-2}$ and $G_8\approx 1444\ GeV^{-8}$ (see \cite{Frasca:2012eq}). This yields, for the cut-off $\Lambda=1\ GeV$,
\begin{figure}[H]
  \includegraphics{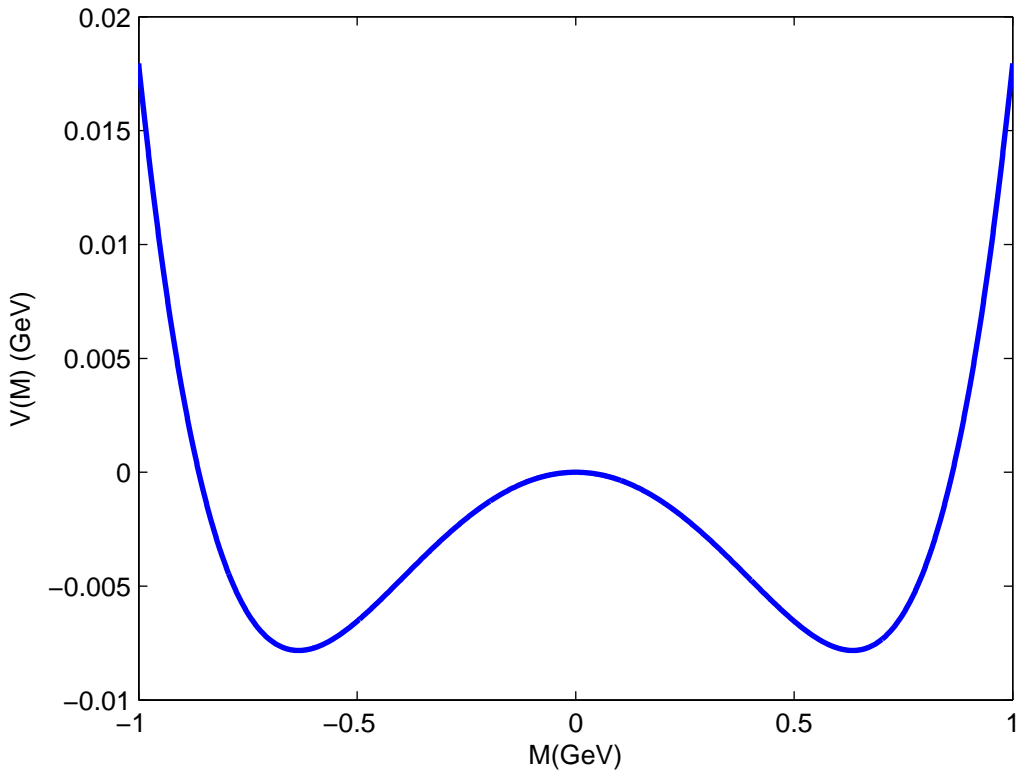}
  \caption{Effective potential for the given values of the coupling constants. This has the form of a mexican hat with a minimum where the chiral phase transition happens.\label{fig:f1}}
\end{figure}
while the first derivative is
\begin{figure}[H]
  \includegraphics{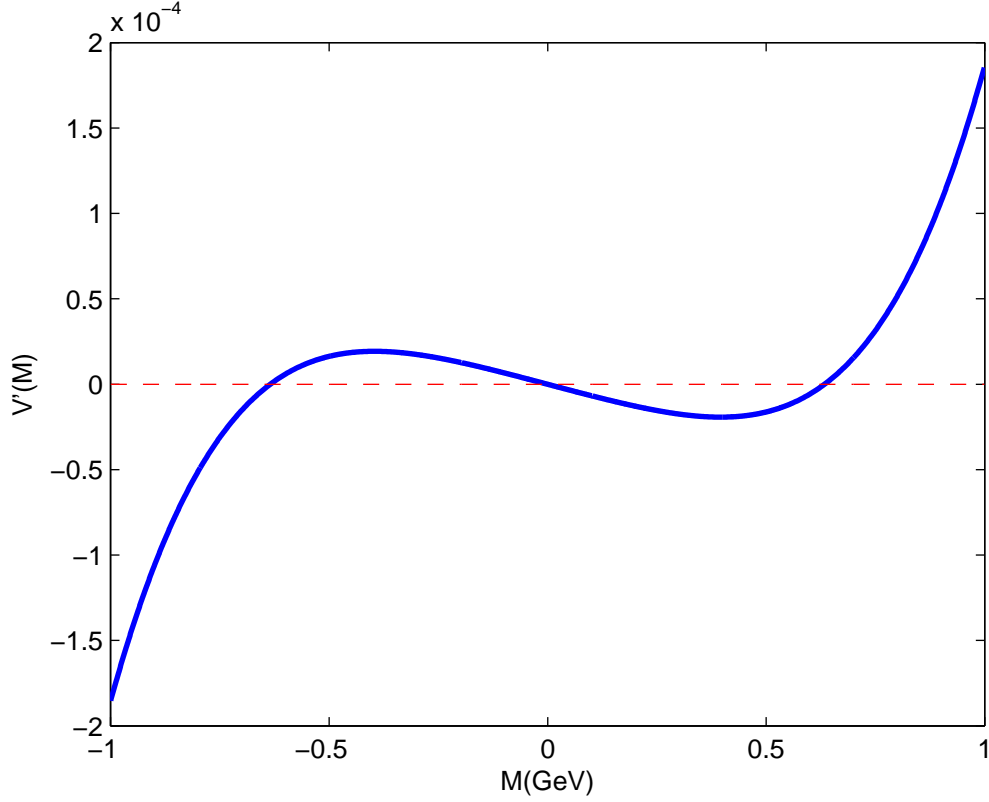}
  \caption{First derivative of the effective potential with respect to the mass $M$. This shows a trivial zero and two non-trivial ones.\label{fig:f2}}.
\end{figure}
The first derivative shows a trivial zero in the origin and two symmetric non-trivial ones. Then, the value of the effective mass for the quarks at the chiral phase transition is $M_{eff}\approx 0.63\ GeV$ for which the effective potential has a minimum. The gap equation is given by \cite{Osipov:2006ns}
\begin{equation}
\label{eq:gape}
    h(M)+\frac{NN_f}{2\pi^2}MJ_0(M^2)=0
\end{equation}
that reduces to the standard one when $G_8=0$. This equation has $M_{eff}=0.63\ GeV$ for a smaller cut-off $\Lambda=0.747\ GeV$ with respect to the case of the effective potential.



\section{Critical temperature\label{sec5}}

In order to compute the critical temperature, we need to evaluate the integral $J_0$ with Matsubara frequencies. For convenience, we set the chemical potential $\mu$ to zero. We have \cite{Hiller:2008nu}
\begin{equation}
\label{eq:J_0T}
    J_0(M^2,T)=J_0(M^2)-8\int_0^\Lambda d|{\bm p}|\frac{|{\bm p}|^2}{\sqrt{|{\bm p}|^2+M^2}}\frac{1}{1+\exp{\beta\sqrt{|{\bm p}|^2+M^2}}}
\end{equation}
being $\beta=1/T$ the inverse of temperature and $J_0(M^2,0)$ that given in eq.(\ref{eq:J_0}). This is completely independent on $G_8$. Then, we can prove the following statement to be true: {\sl in the chiral limit, the critical temperature is unaffected by the eight quark interaction term.} Of course, in real world quarks u and d have different masses and so, this statement is only approximately true.
This can be seen very easily by noting that $\lim_{M\rightarrow 0}h(M)/M=-1/G$, not dependent on $G_8$. This yields the standard equation for $T_c$ of the NJL model, using eq.(\ref{eq:gape}) and eq.(\ref{eq:J_0T}) \cite{GomezDumm:2004sr,Frasca:2011bd}. One has
\begin{equation}
   -\frac{1}{G}+\frac{NN_f}{2\pi^2}J_0(0,T_c)=0.
\end{equation}
A mass gap $m_0$ (see eq.~(\ref{eq:ms})) has the effect to change this equation into \cite{GomezDumm:2004sr,Frasca:2011bd}
\begin{equation}
   -\frac{1}{G_{eff}}+\frac{NN_f}{2\pi^2}J_0(0,T_c)=0
\end{equation}
introducing an effective NJL coupling constant $1/G_{eff}=1/G+m_0^2$. This applies also to the other equations of the model we discussed so far.

\section{Conclusions\label{sec6}}

In this paper we have derived a NJL model and its next-to-leading order correction directly from QCD. We obtained an eight quark interaction term
, starting from a set of exact solutions describing the ground state of the theory.
. This was made possible thanks to the gluon propagator that has
, in this way,
 a closed form. Studies on a Yang-Mills theory without quarks in the Landau gauge, both on the lattice and theoretically, yield strong support to this conclusion. In some way, physicists have obtained a proof of existence of the mass gap in QCD but this cannot be turned yet into a mathematical proof. Surely, it is enough to start to perform computations in QCD and try to obtain in this way analytically calculable observables in the low-energy limit. In this paper we showed that, in the chiral limit with two quarks, the critical temperature of the chiral transition is not influenced by the eight quark interaction term. This appears in this way a leading order approximation to a complete calculation. We hope to present in the near future an extended thermodynamical analysis of the NJl model we obtained and a study of the particle spectrum as well. This should grant a better understanding of the low-energy behavior of QCD.


\begin{acknowledgements}

I am deeply grateful to Orlando Oliveira for providing me the lattice data of the gluon propagator he and his collaborators obtained and that were used in \cite{Duarte:2016iko}.
\end{acknowledgements}


\end{document}